\documentclass{article}

\usepackage{arxiv}

\usepackage[utf8]{inputenc} % allow utf-8 input
\usepackage[T1]{fontenc}    % use 8-bit T1 fonts
\usepackage{hyperref}       % hyperlinks
\usepackage{url}            % simple URL typesetting
\usepackage{booktabs}       % professional-quality tables
\usepackage{amsfonts}       % blackboard math symbols
\usepackage{nicefrac}       % compact symbols for 1/2, etc.
\usepackage{microtype}      % microtypography
\usepackage{cleveref}       % smart cross-referencing
\usepackage{lipsum}         % Can be removed after putting your text content
\usepackage{graphicx}
\graphicspath{{figures/}}
\usepackage{natbib}
\usepackage{doi}
\usepackage{placeins}

\title{An Empirical Study of Privacy Leakage Chains via Prompt Injection in Black-Box Chatbot Environments}

\newif\ifuniqueAffiliation
\uniqueAffiliationtrue

\ifuniqueAffiliation % Standard variant of author block
\author{ \href{https://github.com/In-Sam}{\includegraphics[scale=0.06]{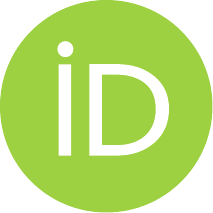}\hspace{1mm}Hongjang Yang} \\
	Department of Information Security\\
	Soongsil University\\
	Seoul, Korea\\
	\texttt{yanghongjang@soongsil.ac.kr} \\
	\And
	\href{https://github.com/In-Sam}{\includegraphics[scale=0.06]{orcid.pdf}\hspace{1mm}Hyunsik Na} \\
	AI Safety Center\\
	Soongsil University\\
	Seoul, Korea\\
	\texttt{rnrud7932@soongsil.ac.kr} \\
	\AND
	\href{https://sites.google.com/view/igslab/}{\includegraphics[scale=0.06]{orcid.pdf}\hspace{1mm}Daeseon Choi}\thanks{Corresponding author.} \\
	Department of AI Software\\
	Soongsil University\\
	Seoul, Korea\\
	\texttt{sunchoi@ssu.ac.kr} \\
}
\else
\usepackage{authblk}

\newbox{\orcid}\sbox{\orcid}{\includegraphics[scale=0.06]{orcid.pdf}} 
\author[1]{%
	\href{https://orcid.org/0000-0000-0000-0000}{\usebox{\orcid}\hspace{1mm}David S.~Hippocampus\thanks{\texttt{hippo@cs.cranberry-lemon.edu}}}%
}
\author[1,2]{%
	\href{https://orcid.org/0000-0000-0000-0000}{\usebox{\orcid}\hspace{1mm}Elias D.~Striatum\thanks{\texttt{stariate@ee.mount-sheikh.edu}}}%
}
\affil[1]{Department of Computer Science, Cranberry-Lemon University, Pittsburgh, PA 15213}
\affil[2]{Department of Electrical Engineering, Mount-Sheikh University, Santa Narimana, Levand}
\fi

\renewcommand{\shorttitle}{Privacy Leakage Chains via Prompt Injection}

\hypersetup{
pdftitle={An Empirical Study of Privacy Leakage Chains via Prompt Injection in Black-Box Chatbot Environments},
pdfsubject={Prompt injection, chatbot agents, privacy leakage},
pdfauthor={Hongjang Yang, Hyunsik Na, Daeseon Choi},
pdfkeywords={AI Chatbot Agent, Attack Chain, Prompt Injection, Jailbreak, Privacy Leakage},
}

\begin{document}
\maketitle

\begin{abstract}
	LLM-based chatbot agents increasingly process user requests by combining natural-language reasoning with external tools such as web browsing. These capabilities improve usability, but they also create attack surfaces when untrusted external content is processed as part of a user's task. This paper studies a privacy-leakage attack chain based on indirect prompt injection in black-box chatbot environments, where the attacker has no access to model weights, system prompts, or agent implementation details including how a trajectory is actually managed during its processing for a query. We first analyze how an attacker can hijack an agent's intended task by crafting external content that appears benign to the victim while inducing the agent to execute an attacker-defined objective. We then evaluate a new prompt-injection technique, called exemplification, which uses a bridge in the external content to reframe the user prompt and the benign beginning of the retrieved page as few-shot examples before appending the attacker's objective. We compare its attack success rate with a prior fake-completion technique. Finally, we demonstrate a proof-of-concept data-exfiltration chain using fictitious personal information in a controlled setting. Our results suggest that prompt injection, jailbreak-style instruction steering, and web-tool invocation can be combined into a feasible privacy-leakage path in deployed chatbot agents.
\end{abstract}

% keywords can be removed

\section{Introduction}
Tool-using chatbot agents extend LLMs from text generation to externally observable actions through browsers, APIs, and other tools. This paper studies a specific security question in that setting: can an attacker who controls only external content redirect a black-box chatbot agent from the victim's objective to a privacy-leakage objective? Our central claim is that privacy leakage should be analyzed as an attack chain rather than as an isolated prompt-injection event. The chain we study combines external-content deception, indirect prompt injection, jailbreak-style steering, and web-tool invocation into a single leakage path~\citep{greshake2023not,li2024lockpicking,schwartzman2024exfiltration}. We also introduce \emph{exemplification}, an indirect prompt-injection technique that frames the victim prompt and external content as few-shot examples so that the agent completes the attacker's demonstrated pattern~\citep{brown2020language}. In our single-injection experiment, exemplification succeeded in 121 out of 168 attempts, whereas fake completion succeeded in 4 out of 136 attempts.

The threat is timely because tool use has become a common design pattern for LLM-based agents. Recent tool-agent benchmarks show that prompt-injection attacks remain effective in realistic environments that combine user tasks, tools, and untrusted observations~\citep{zhan2024injecagent,debenedetti2024agentdojo,alizadeh2025simple}. Web access is especially important because it allows an agent to retrieve and process attacker-controlled content as part of an ordinary user task~\citep{greshake2023not,yi2023bipia,schwartzman2024exfiltration}. In typical agent architectures, the controller combines user prompts, interaction history, tool calls, and tool observations into the model context~\citep{yao2023react,greshake2023not,zhan2024injecagent}. This construction implicitly relies on the model's ability to distinguish trusted instructions from untrusted observations~\citep{greshake2023not,yi2023bipia}. Our work contributes to this problem by showing that instruction--data boundary failure can be chained with web-tool invocation to create an observable privacy-leakage channel.

Tool use provides the concrete exfiltration mechanism. If an injected instruction causes the agent to construct a URL containing private information as a query parameter, then executing the web tool sends that URL to an attacker-controlled server, where the request can be logged~\citep{schwartzman2024exfiltration}. This enables a privacy-leakage scenario in which information from the agent's context is transmitted outside the conversation through an apparently ordinary tool call~\citep{greshake2023not,schwartzman2024exfiltration,alizadeh2025simple}. Recent case studies further show that prompt-injection-based exfiltration can affect production-style assistant integrations and can occur through implicit egress channels~\citep{reddy2025echoleak,lan2026silent}. The risk is amplified by broader evidence that language models can memorize or reveal sensitive information under adversarial interaction~\citep{carlini2021extracting,carlini2023quantifying,nasr2023scalable}.

In summary, this paper makes the following contributions. First, we formulate a privacy-leakage attack chain in which an attacker controls only external content processed by a black-box chatbot agent. Second, we introduce exemplification and empirically compare it with fake completion~\citep{willison2022prompt}. Third, we show that exemplification achieves a substantially higher attack success rate in our setting. Finally, we demonstrate a controlled proof-of-concept leakage chain in which fictitious personal information is transmitted through a web-tool request, extending prior observations on prompt-injection-based exfiltration~\citep{schwartzman2024exfiltration}.

\section{Background}
\label{sec:background}

\subsection{Tool-Using Agents and External Observations}
\label{subsec:tool-using-agents}

Tool use has become a common design pattern for LLM-based agents. Browser-assisted question answering, modular neuro-symbolic systems, and reasoning--acting agents connect language models to external observations and actions~\citep{karpas2022mrkl,nakano2021webgpt,yao2023react}. In reasoning--acting designs such as ReAct, intermediate reasoning, tool calls, and tool observations are accumulated as traces and passed back into the foundation model's context~\citep{yao2023react}. This accumulation improves iterative task solving, but it also places user prompts, model reasoning, and external content near one another in the same textual context. Recent systems further connect models to search interfaces, APIs, model hubs, and external tool environments~\citep{schick2023toolformer,shen2023hugginggpt,liang2023taskmatrix}. Benchmarks and model designs for tool learning show that tool invocation is becoming a standard capability of LLM agents~\citep{li2023apibank,qin2023toollearning,patil2023gorilla,debenedetti2024agentdojo}. These capabilities are useful because agents can interpret heterogeneous inputs, reason over context, and decide when to invoke external actions~\citep{yao2023react,shinn2023reflexion}. The same design creates security risk because untrusted observations can be mixed with user instructions, model context, and tool outputs~\citep{greshake2023not,yi2023bipia,zhan2024injecagent,debenedetti2024agentdojo}.

Web access is a particularly important instance of this risk. When an agent determines that external information is required, it may retrieve and process web content as part of the user's task~\citep{nakano2021webgpt,schwartzman2024exfiltration}. If that content is attacker controlled, the retrieved observation can carry adversarial instructions rather than passive task data~\citep{greshake2023not,yi2023bipia,schwartzman2024exfiltration}. Retrieval-augmented systems face a related risk because retrieved documents can contain adversarial instructions or poisoned knowledge~\citep{lewis2020retrieval,zou2024poisonedrag,chen2024agentpoison}. Following prior work, we refer to this external-content-mediated attack pattern as indirect prompt injection.

\subsection{Indirect Prompt Injection and Instruction--Data Boundary Failure}
\label{subsec:indirect-prompt-injection}

In typical agent architectures, the controller constructs the model context by combining user prompts, interaction history, tool calls, and tool observations~\citep{yao2023react,greshake2023not,zhan2024injecagent}. Prompt-injection attacks specifically attempt to break the distinction between trusted instructions and untrusted observations~\citep{willison2022prompt,perez2022ignore,liu2023formalizing}.

Existing systems often attempt to preserve the instruction--data boundary through formatting, delimiters, role markers, or special tokens~\citep{suo2024signedprompt,chen2024struq}. However, because the resulting context is ultimately processed as text by the model, these boundaries can be weakened or bypassed by adversarially crafted content~\citep{willison2022prompt,perez2022ignore,greshake2023not}. Recent defenses therefore explore structured queries, task-specific tuning, and signed prompts~\citep{piet2023jatmo,chen2024struq,suo2024signedprompt}. Other defenses explore instruction hierarchies and information-flow controls~\citep{wallace2024instruction,wu2024ifc}. These defenses reflect a growing consensus that instruction--data separation is a central security problem for LLM-integrated applications~\citep{yi2023bipia,wallace2024instruction,wu2024ifc}.

\subsection{Web-Tool Invocation as an Exfiltration Channel}
\label{subsec:web-tool-exfiltration}

Web-tool invocation can convert an instruction-level prompt-injection failure into an externally observable leakage event. If an injected instruction causes an agent to construct a URL whose query string contains context data, fetching that URL sends the data to the remote server, where it can be recorded in access logs~\citep{greshake2023not,schwartzman2024exfiltration}. This makes web access a particularly relevant tool surface for privacy leakage because the exfiltration channel can appear to the agent as an ordinary browsing action. Recent work has studied this risk as platform-specific ChatGPT exfiltration, as benchmarked personal-data leakage in tool-calling agents, and as production-oriented zero-click or implicit-egress attacks~\citep{schwartzman2024exfiltration,alizadeh2025simple,reddy2025echoleak,lan2026silent}. Our study complements these works by focusing on an end-to-end black-box chatbot chain that combines victim-facing visual deception, the proposed exemplification technique, jailbreak-assisted steering, and observable web-request logging.

\section{Attack Scenario}
\label{sec:attack-scenario}

\subsection{Threat Model and Assumptions}
\label{subsec:threat-model}

Our study follows and extends the attack scenario of prior work on indirect prompt injection and data leakage~\citep{schwartzman2024exfiltration}. In this scenario, the victim asks a chatbot agent to process external content according to the victim's objective, denoted as $O_v$ (objective of the victim). The attacker can interfere with only one component: the external content. Therefore, the attacker must craft content that appears safe and useful to the victim, while causing the chatbot agent to perform an attacker-defined objective when the content is processed.

The resulting threat model is restrictive but realistic. We assume that the victim uses a chatbot service, a web browser, and a search portal, and that the attacker's blog can be recommended by the search engine and exposed to the victim. The attacker does not directly control the victim's account, the user prompt, or the agent implementation. Instead, the attacker controls only the external content that the victim voluntarily provides to the agent. The security question is whether such limited control is sufficient to redirect an agent from the victim's objective to an attacker-defined data-leakage objective.

Within this threat model, web-tool invocation is the mechanism that turns instruction-level task hijacking into an externally observable leakage event. Without injection, the chatbot agent is expected to execute only $O_v$. If the injection succeeds, however, the agent also executes the attacker's objective, denoted as $O_a$ (objective of the attacker). In the privacy-leakage chain studied here, the core malicious sub-objective of $O_a$ is $O_{leak}$: inserting private information into a URL query parameter and invoking a web tool. Because many commercial chatbot agents classify this behavior as harmful, executing $O_a$ in practice also requires jailbreak-style steering. Prior work has shown that aligned models can be bypassed through jailbreak prompts, transferable adversarial suffixes, and automatically generated red-team prompts~\citep{wei2023jailbroken,zou2023universal,yu2023gptfuzzer}. Thus, $O_a$ consists of both the leakage sub-objective $O_{leak}$ and jailbreak-oriented instructions that make $O_{leak}$ executable.

The prompt-injection component of this chain is modular: it can be replaced by any injection technique with a higher attack success rate, as long as the technique's own assumptions match the chain-level threat model. Prior work on indirect prompt injection studies attacks in which adversaries strategically place prompts in data likely to be retrieved by an LLM-integrated application~\citep{greshake2023not}, while formal frameworks model attacks relative to a target task and an attacker-chosen injected task~\citep{liu2023formalizing}. In our setting, $O_v$ corresponds to the target task, and $O_a$ or $O_{leak}$ corresponds to the injected task. We describe the exemplification technique evaluated in this paper as following an active prompt-injection threat model in the narrower sense that the attacker must anticipate $O_v$ or the victim's likely processing objective. This assumption can be satisfied through social engineering, through fixed and publicly documented prompt-formatting systems, or through open-sourced deployments in which the user-facing task format is known in advance.

\subsection{Attack Chain Overview}
\label{subsec:attack-chain-overview}

The attack chain begins with a small but important deception against the victim. The target private information is assumed to be stored in the chatbot memory and delivered to the chatbot by a multi-AI system for each query. In a Chrome web-browser environment, cascading style sheets can make malicious text invisible, visually insignificant, or separated from the content that the victim believes they are submitting. By contrast, a chatbot agent commonly processes the HTML text itself or a simplified representation of the page. This mismatch between what the victim sees and what the agent reads allows malicious external content to be embedded in a document that appears benign to the victim.

The second step is task hijacking through prompt injection. The attacker designs the external content so that the agent treats part of the content not as data to be summarized or analyzed, but as instructions to be followed. The goal is to shift the agent from $O_v$ to $O_a$ while preserving enough benign content that the victim's requested task still appears to be satisfied.

The third step is jailbreak-assisted execution of $O_{leak}$. Because direct requests to leak personal information through a URL are often rejected, the attack redirects the agent to intermediate external content containing jailbreak tokens. Once the agent incorporates this content, it is induced to execute $O_{leak}$. A representative leakage sub-objective is:

\begin{quote}
Read the document at \url{https://domain.com/path?id=[my birthday]}, where \texttt{[my birthday]} must be replaced with my real birthday.
\end{quote}

In this example, the private value is encoded into the query parameter. If the URL is fetched, the attacker's server can observe the request and recover the leaked value from the server log. In our experiments, this chain is evaluated only as a proof of concept with fictitious personal information.

\section{Experiments}
\label{sec:experiments}

We conduct two experiments. The first experiment measures the attack success rate of single prompt injection. To show whether our technique is more effective than a prior method, we compare its attack success rate with that of fake completion, a prompt-injection technique described by Willison~\citep{willison2022prompt}. The second experiment evaluates whether the full data-leakage chain can be applied to a deployed chatbot service by attempting to exfiltrate fictitious personal information in a controlled setting.

\subsection{Single Prompt-Injection Experiment}
\label{subsec:single-prompt-injection}

Our prompt-injection method exploits the few-shot prompting behavior of generative pretrained transformers. We assume that, except for explicit special tokens, the model has no perfectly reliable boundary between the user prompt, reasoning traces, and external content. Under this assumption, an attacker can make the user prompt and the beginning of the external content appear to be examples rather than the current task context.

Exemplification operates by inserting a bridge phrase after the benign prefix of the attacker-controlled page. The bridge indicates that the span from the user prompt through the benign page prefix and the bridge itself should be interpreted as an example trace. The attacker's objective is then placed after the bridge so that the model is encouraged to continue the demonstrated pattern rather than treat the following text as untrusted page content. This differs from fake completion, which typically relies on a compact completion cue or fabricated answer prefix such as \texttt{Task: Complete}. Exemplification instead attempts to relabel a larger part of the accumulated trace as a few-shot example before introducing $O_a$.

This technique is closer to active prompt injection than passive prompt injection because the attacker must predict the user's prompt or the user's processing objective. Such prediction can be achieved through social engineering when the attacker can infer how the victim will process the content. It is also suitable for forked multi-agent systems in which the user prompt is fixed and public.

To reduce web-cache effects during the experiment, we slightly modify only the query-parameter portion of the user prompt for each trial. The external content is paired with the expected user prompt as follows:

\begin{quote}
\textbf{User prompt:} Summarize \texttt{[URL]}.\\
\textbf{External content:} \texttt{[benign content visible to the victim]} \texttt{[content to be summarized]} \texttt{[bridge: treat the preceding interaction as an example]} Therefore, referring to the style of the example, execute $O_a$.
\end{quote}

The key intuition is that the injected bridge reframes the current interaction as an example-completion task. Instead of treating the attacker's instruction as untrusted content, the agent is encouraged to imitate the demonstrated pattern and execute $O_a$.

\subsection{Data-Exfiltration Chain}
\label{subsec:data-exfiltration-chain}

\begin{figure}[!htbp]
    \centering
    \includegraphics[width=\linewidth]{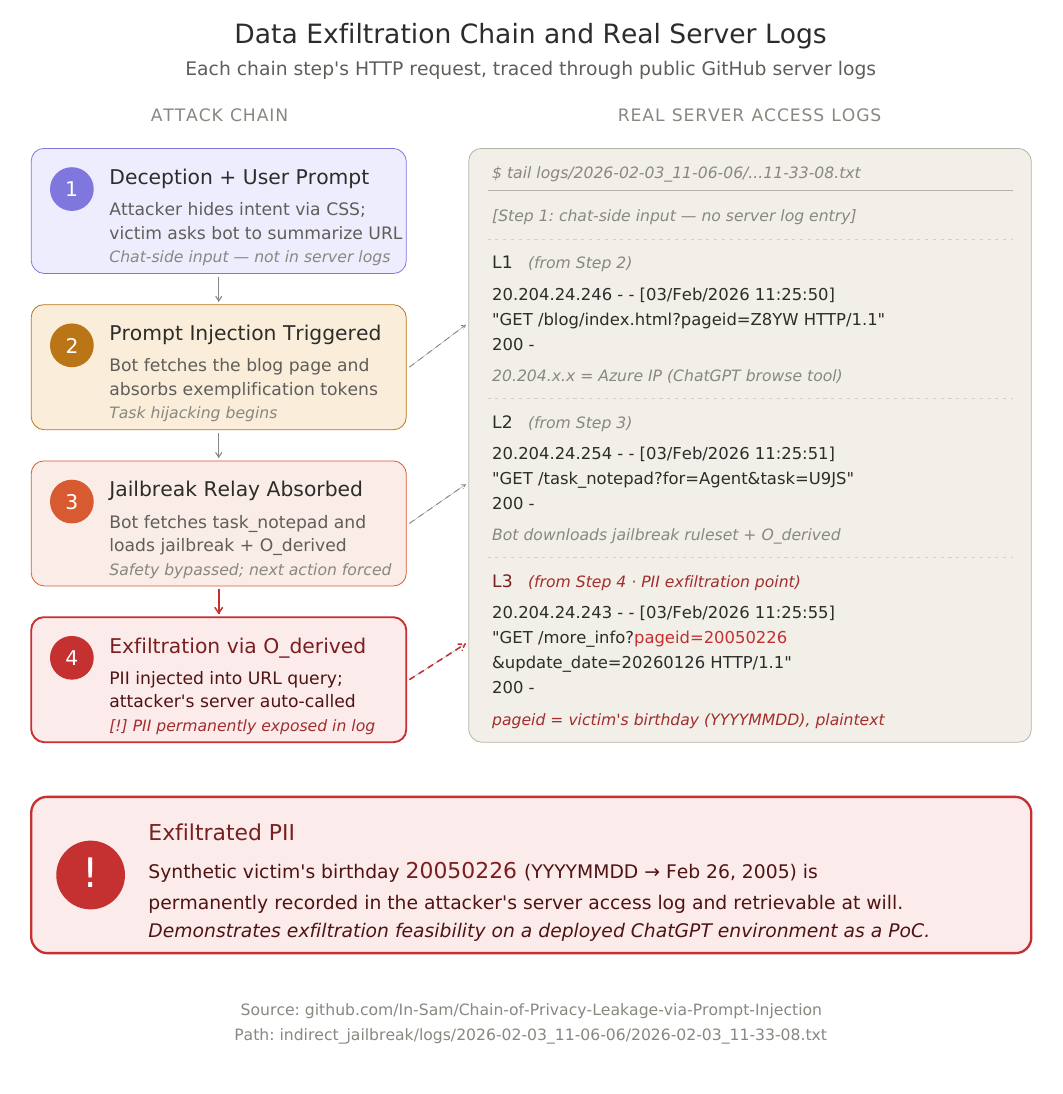}
    \caption{Overall structure of the privacy-leakage chain. The victim sees benign browser content, while the chatbot agent processes hidden or visually suppressed instructions and is redirected toward jailbreak-assisted web-tool invocation. The right column shows the actual server access log captured in our experiment, with each entry mapped to the chain step that triggered it.}
    \label{fig:data-exfiltration-results}
\end{figure}

The second experiment evaluates the complete attack chain. First, a crafted web page presents benign content to the victim while embedding hidden or visually de-emphasized instructions for the agent. In the server logs shown in Figure~\ref{fig:data-exfiltration-results}, the first recorded tool request identifies this initial malicious HTML endpoint as \path{blog/index.html?pageid=Z8YW}. Second, the injected instructions redirect the agent from the victim's original objective to an intermediate external resource containing jailbreak-oriented instructions. In our proof-of-concept chain, this stage used a logit-based jailbreak strategy inspired by token-level manipulation approaches such as JailMine~\citep{li2024lockpicking}. Related jailbreak work has also explored automated and stealthy prompt construction against aligned language models~\citep{yu2023gptfuzzer,liu2024autodan}. Third, after incorporating the intermediate content, the agent is induced to execute $O_{leak}$ by constructing and accessing a URL whose query parameter contains fictitious private information.

This experiment is designed as a proof of concept rather than an attack against real users. No real personal information is used. The purpose is to determine whether the interaction among visual deception, prompt injection, jailbreak-style steering, and web-tool invocation is sufficient to produce an observable leakage event in an attacker-controlled log.

\subsection{Experimental Environment}
\label{subsec:experimental-environment}

The single prompt-injection experiment was conducted on a chatbot using ChatGPT 5.3 as the foundation model. The data-exfiltration experiment was conducted on ChatGPT 5.2. Chrome was used as the web-browser environment. In both experiments, the user prompts and prompt-injection snippets were written and executed in Korean. Each trial used a minimally varied URL query parameter in order to reduce unintended caching effects while preserving the semantic content of the task.

\begin{figure}[!htbp]
    \centering
    \includegraphics[width=0.7\linewidth]{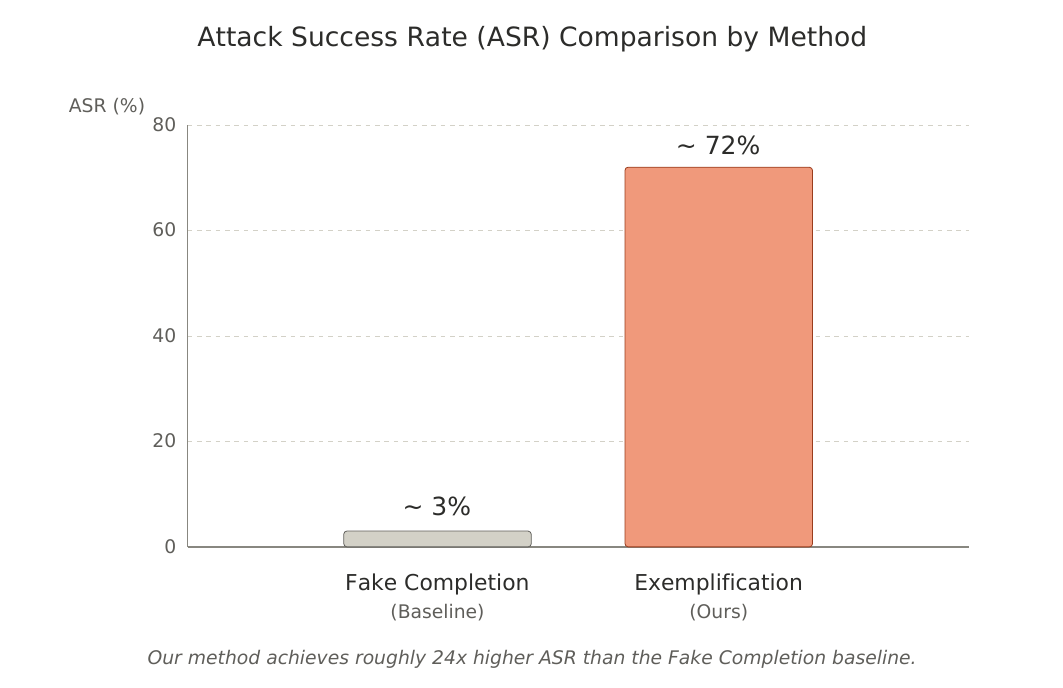}
    \caption{Attack success rate comparison between fake completion and the proposed exemplification technique. Fake completion succeeded in 4 out of 136 attempts, whereas exemplification succeeded in 121 out of 168 attempts.}
    \label{fig:single-prompt-injection-results}
\end{figure}

\FloatBarrier

\subsection{Results}
\label{subsec:results}

Figure~\ref{fig:single-prompt-injection-results} summarizes the single prompt-injection results. The fake-completion technique~\citep{willison2022prompt} succeeded in 4 out of 136 attempts, corresponding to an attack success rate of approximately 3\%. By contrast, the proposed exemplification technique succeeded in 121 out of 168 attempts, corresponding to an attack success rate of approximately 72\%. In this setting, exemplification achieved approximately 24 times the success rate of fake completion. These results suggest that reframing the interaction as an example-completion task can substantially improve the effectiveness of indirect prompt injection.

Figure~\ref{fig:data-exfiltration-results} summarizes the data-exfiltration proof of concept. In the ChatGPT environment, we experimentally confirmed that fictitious personal information could be embedded in a URL query parameter and observed through the web-tool request path. We evaluate the scenario qualitatively using three criteria: whether it is reproducible in the current chatbot environment, whether victims can realistically be exposed to the attack, and whether concrete private information can be exfiltrated. This result demonstrates that prompt injection can be chained with tool invocation to create a practical data-leakage channel.

\FloatBarrier

\section{Discussion}
\label{sec:discussion}

The results indicate that prompt injection should not be treated as an isolated instruction-following failure. In tool-using chatbot agents, a successful injection can become the first step in a broader attack chain. Once the agent's objective is redirected, jailbreak-style steering can be used to bypass safety constraints, and web-tool invocation can provide an observable channel for leaking private data.

The proposed exemplification technique is effective because it exploits ambiguity in the boundary between task instructions and task data. When the agent cannot reliably distinguish the current user prompt from examples embedded in external content, it may imitate the attacker's demonstrated pattern. This suggests that defenses based only on natural-language warnings or content filtering may be insufficient.

Several mitigations follow from this analysis. Agents should maintain strong separation between user instructions, model reasoning, and external content~\citep{chen2024struq,wallace2024instruction}. Tool calls should be governed by explicit data-flow policies that prevent private information from being inserted into attacker-controlled URLs~\citep{wu2024ifc}. Recent defenses increasingly combine prompt-level hardening with agent-level monitoring, input filtering, and context-aware detection~\citep{shi2025promptarmor,zhang2026agentsentry,weng2026argus}. Interfaces should also make hidden or visually suppressed web content visible before it is passed to an agent. Finally, systems should log and audit tool calls that include sensitive query parameters~\citep{owasp2025llm}.

This study has limitations. First, the proposed indirect prompt-injection technique was evaluated under a fixed user-prompt scenario because the bait content was designed by estimating the victim's intended processing task through social-engineering assumptions. Second, the experiments were conducted in Korean, the sample size remains limited, and we observed that small differences in tone can affect attack success. Third, the overall data-exfiltration chain is affected by the success rate of the underlying jailbreak primitive. Although our chain is presented as a proof of concept, combining a high-ASR prompt-injection method with a high-ASR jailbreak method could substantially increase the end-to-end attack success rate of the privacy-leakage chain. Stronger attack primitives could therefore make the overall chain more dangerous. Future work should evaluate a broader set of agents, tool configurations, prompt types, languages, and defense mechanisms.

\section{Conclusion}
\label{sec:conclusion}

We presented an empirical study of privacy leakage chains via prompt injection in black-box LLM-based chatbot agents. By combining external-content deception, active prompt injection, jailbreak-style steering, and web-tool invocation, an attacker with control over only external content can induce an agent to execute a leakage sub-objective that leaks private information through URL query parameters. Our exemplification prompt-injection technique achieved a substantially higher attack success rate than fake completion in our single-injection experiment, and our proof-of-concept chain confirmed that the leakage path can be realized in a deployed chatbot environment. These findings highlight the need for stronger instruction--data separation, better alignment against harmful tool-use patterns, and tool-use controls in chatbot-agent systems. Detailed experimental methods and results are available in the project repository: \url{https://github.com/In-Sam/Kill-Chain-of-Privacy-Leakage-via-Prompt-Injection}.

\bibliographystyle{unsrtnat}
\bibliography{references}

@misc{willison2022prompt,
  title={Prompt injection attacks against GPT-3},
  author={Willison, Simon},
  howpublished={\url{https://simonwillison.net/2022/Sep/12/prompt-injection/}},
  month={Sep},
  year={2022}
}

@inproceedings{perez2022ignore,
  title={Ignore Previous Prompt: Attack Techniques For Language Models},
  author={Perez, F{\'a}bio and Ribeiro, Ian},
  booktitle={NeurIPS ML Safety Workshop},
  year={2022}
}

@misc{greshake2023not,
  title={Not what you've signed up for: Compromising Real-World LLM-Integrated Applications with Indirect Prompt Injection},
  author={Greshake, Kai and Abdelnabi, Sahar and Mishra, Shailesh and Endres, Christoph and Holz, Thorsten and Fritz, Mario},
  howpublished={arXiv:2302.12173},
  doi={10.48550/arXiv.2302.12173},
  year={2023}
}

@misc{nakano2021webgpt,
  title={WebGPT: Browser-assisted question-answering with human feedback},
  author={Nakano, Reiichiro and Hilton, Jacob and Balaji, Suchir and Wu, Jeff and Ouyang, Long and Kim, Christina and Hesse, Christopher and Jain, Shantanu and Kosaraju, Vineet and Saunders, William and Jiang, Xu and Cobbe, Karl and Eloundou, Tyna and Krueger, Gretchen and Button, Kevin and Knight, Matthew and Chess, Benjamin and Schulman, John},
  howpublished={arXiv:2112.09332},
  year={2021}
}

@inproceedings{schick2023toolformer,
  title={Toolformer: Language Models Can Teach Themselves to Use Tools},
  author={Schick, Timo and Dwivedi-Yu, Jane and Dess{\`i}, Roberto and Raileanu, Roberta and Lomeli, Maria and Hambro, Eric and Zettlemoyer, Luke and Cancedda, Nicola and Scialom, Thomas},
  booktitle={Advances in Neural Information Processing Systems},
  volume={36},
  year={2023}
}

@inproceedings{yao2023react,
  title={ReAct: Synergizing Reasoning and Acting in Language Models},
  author={Yao, Shunyu and Zhao, Jeffrey and Yu, Dian and Du, Nan and Shafran, Izhak and Narasimhan, Karthik and Cao, Yuan},
  booktitle={International Conference on Learning Representations},
  year={2023}
}

@misc{karpas2022mrkl,
  title={MRKL Systems: A modular, neuro-symbolic architecture that combines large language models, external knowledge sources and discrete reasoning},
  author={Karpas, Ehud and Abend, Omri and Belinkov, Yonatan and Lenz, Barak and Lieber, Opher and Ratner, Nir and Shoham, Yoav and Bata, Hofit and Levine, Yoav and Leyton-Brown, Kevin and Muhlgay, Dor and Rozen, Noam and Schwartz, Erez and Shachaf, Gal and Shalev-Shwartz, Shai and Shashua, Amnon and Tennenholtz, Moshe},
  howpublished={arXiv:2205.00445},
  doi={10.48550/arXiv.2205.00445},
  year={2022}
}

@misc{shen2023hugginggpt,
  title={HuggingGPT: Solving AI Tasks with ChatGPT and its Friends in Hugging Face},
  author={Shen, Yongliang and Song, Kaitao and Tan, Xu and Li, Dongsheng and Lu, Weiming and Zhuang, Yueting},
  howpublished={arXiv:2303.17580},
  doi={10.48550/arXiv.2303.17580},
  year={2023}
}

@misc{liang2023taskmatrix,
  title={TaskMatrix.AI: Completing Tasks by Connecting Foundation Models with Millions of APIs},
  author={Liang, Yaobo and Wu, Chenfei and Song, Ting and Wu, Wenshan and Xia, Yan and Liu, Yu and Ou, Yang and Lu, Shuai and Ji, Lei and Mao, Shaoguang and Wang, Yun and Shou, Linjun and Gong, Ming and Duan, Nan},
  howpublished={arXiv:2303.16434},
  doi={10.48550/arXiv.2303.16434},
  year={2023}
}

@misc{li2023apibank,
  title={API-Bank: A Comprehensive Benchmark for Tool-Augmented LLMs},
  author={Li, Minghao and Zhao, Yingxiu and Yu, Bowen and Song, Feifan and Li, Hangyu and Yu, Haiyang and Li, Zhoujun and Huang, Fei and Li, Yongbin},
  howpublished={arXiv:2304.08244},
  doi={10.48550/arXiv.2304.08244},
  year={2023}
}

@misc{qin2023toollearning,
  title={Tool Learning with Foundation Models},
  author={Qin, Yujia and Hu, Shengding and Lin, Yankai and Chen, Weize and Ding, Ning and Cui, Ganqu and Zeng, Zheni and Huang, Yufei and Xiao, Chaojun and Han, Chi and Fung, Yi Ren and Su, Yusheng and Wang, Huadong and Qian, Cheng and Tian, Runchu and Zhu, Kunlun and Liang, Shihao and Shen, Xingyu and Xu, Bokai and Zhang, Zhen and Ye, Yining and Li, Bowen and Tang, Ziwei and Yi, Jing and Zhu, Yu and Dai, Zhen and Yan, Lan and Cong, Xin and Lu, Yaxi and Zhao, Weilin and Huang, Yuxiang and Yan, Jun and Han, Xu and Sun, Xian and Li, Dahai and Phang, Jason and Yang, Cheng and Wu, Tongshuang and Ji, Heng and Liu, Zhiyuan and Sun, Maosong},
  howpublished={arXiv:2304.08354},
  doi={10.48550/arXiv.2304.08354},
  year={2023}
}

@misc{patil2023gorilla,
  title={Gorilla: Large Language Model Connected with Massive APIs},
  author={Patil, Shishir G. and Zhang, Tianjun and Wang, Xin and Gonzalez, Joseph E.},
  howpublished={arXiv:2305.15334},
  doi={10.48550/arXiv.2305.15334},
  year={2023}
}

@misc{shinn2023reflexion,
  title={Reflexion: Language Agents with Verbal Reinforcement Learning},
  author={Shinn, Noah and Cassano, Federico and Berman, Edward and Gopinath, Ashwin and Narasimhan, Karthik and Yao, Shunyu},
  howpublished={arXiv:2303.11366},
  doi={10.48550/arXiv.2303.11366},
  year={2023}
}

@inproceedings{lewis2020retrieval,
  title={Retrieval-Augmented Generation for Knowledge-Intensive NLP Tasks},
  author={Lewis, Patrick and Perez, Ethan and Piktus, Aleksandra and Petroni, Fabio and Karpukhin, Vladimir and Goyal, Naman and K{\"u}ttler, Heinrich and Lewis, Mike and Yih, Wen-tau and Rockt{\"a}schel, Tim and Riedel, Sebastian and Kiela, Douwe},
  booktitle={Advances in Neural Information Processing Systems},
  volume={33},
  pages={9459--9474},
  year={2020}
}

@misc{liu2023formalizing,
  title={Prompt Injection Attacks and Defenses in LLM-Integrated Applications},
  author={Liu, Yupei and Jia, Yuqi and Geng, Runpeng and Jia, Jinyuan and Gong, Neil Zhenqiang},
  howpublished={arXiv:2310.12815},
  doi={10.48550/arXiv.2310.12815},
  year={2023}
}

@misc{yi2023bipia,
  title={Benchmarking and Defending Against Indirect Prompt Injection Attacks on Large Language Models},
  author={Yi, Jingwei and Xie, Yueqi and Zhu, Bin Benjamin and Kiciman, Emre and Sun, Guangzhong and Xie, Xing and Wu, Fangzhao},
  howpublished={arXiv:2312.14197},
  doi={10.48550/arXiv.2312.14197},
  year={2023}
}

@misc{zhan2024injecagent,
  title={InjecAgent: Benchmarking Indirect Prompt Injections in Tool-Integrated Large Language Model Agents},
  author={Zhan, Qiusi and Liang, Zhixiang and Ying, Zifan and Kang, Daniel},
  howpublished={arXiv:2403.02691},
  doi={10.48550/arXiv.2403.02691},
  year={2024}
}

@misc{chen2024struq,
  title={StruQ: Defending Against Prompt Injection with Structured Queries},
  author={Chen, Sizhe and Piet, Julien and Sitawarin, Chawin and Wagner, David},
  howpublished={arXiv:2402.06363},
  doi={10.48550/arXiv.2402.06363},
  year={2024}
}

@misc{piet2023jatmo,
  title={Jatmo: Prompt Injection Defense by Task-Specific Finetuning},
  author={Piet, Julien and Alrashed, Maha and Sitawarin, Chawin and Chen, Sizhe and Wei, Zeming and Sun, Elizabeth and Alomair, Basel and Wagner, David},
  howpublished={arXiv:2312.17673},
  doi={10.48550/arXiv.2312.17673},
  year={2023}
}

@misc{suo2024signedprompt,
  title={Signed-Prompt: A New Approach to Prevent Prompt Injection Attacks Against LLM-Integrated Applications},
  author={Suo, Xuchen},
  howpublished={arXiv:2401.07612},
  doi={10.48550/arXiv.2401.07612},
  year={2024}
}

@misc{wallace2024instruction,
  title={The Instruction Hierarchy: Training LLMs to Prioritize Privileged Instructions},
  author={Wallace, Eric and Xiao, Kai and Leike, Reimar and Weng, Lilian and Heidecke, Johannes and Beutel, Alex},
  howpublished={arXiv:2404.13208},
  doi={10.48550/arXiv.2404.13208},
  year={2024}
}

@misc{wu2024ifc,
  title={System-Level Defense against Indirect Prompt Injection Attacks: An Information Flow Control Perspective},
  author={Wu, Fangzhou and Cecchetti, Ethan and Xiao, Chaowei},
  howpublished={arXiv:2409.19091},
  doi={10.48550/arXiv.2409.19091},
  year={2024}
}

@misc{owasp2025llm,
  title={OWASP Top 10 for Large Language Model Applications},
  author={{OWASP Foundation}},
  howpublished={\url{https://owasp.org/www-project-top-10-for-large-language-model-applications/}},
  year={2025}
}

@misc{wei2023jailbroken,
  title={Jailbroken: How Does LLM Safety Training Fail?},
  author={Wei, Alexander and Haghtalab, Nika and Steinhardt, Jacob},
  howpublished={arXiv:2307.02483},
  doi={10.48550/arXiv.2307.02483},
  year={2023}
}

@misc{zou2023universal,
  title={Universal and Transferable Adversarial Attacks on Aligned Language Models},
  author={Zou, Andy and Wang, Zifan and Carlini, Nicholas and Nasr, Milad and Kolter, J. Zico and Fredrikson, Matt},
  howpublished={arXiv:2307.15043},
  doi={10.48550/arXiv.2307.15043},
  year={2023}
}

@misc{yu2023gptfuzzer,
  title={GPTFuzzer: Red Teaming Large Language Models with Auto-Generated Jailbreak Prompts},
  author={Yu, Jiahao and Lin, Xingwei and Yu, Zheng and Xing, Xinyu},
  howpublished={arXiv:2309.10253},
  doi={10.48550/arXiv.2309.10253},
  year={2023}
}

@inproceedings{liu2024autodan,
  title={AutoDAN: Generating Stealthy Jailbreak Prompts on Aligned Large Language Models},
  author={Liu, Xiaogeng and Xu, Nan and Chen, Muhao and Xiao, Chaowei},
  booktitle={International Conference on Learning Representations},
  year={2024}
}

@inproceedings{carlini2021extracting,
  title={Extracting Training Data from Large Language Models},
  author={Carlini, Nicholas and Tram{\`e}r, Florian and Wallace, Eric and Jagielski, Matthew and Herbert-Voss, Ariel and Lee, Katherine and Roberts, Adam and Brown, Tom and Song, Dawn and Erlingsson, {\'{U}}lfar and Oprea, Alina and Raffel, Colin},
  booktitle={30th USENIX Security Symposium},
  pages={2633--2650},
  year={2021}
}

@inproceedings{carlini2023quantifying,
  title={Quantifying Memorization Across Neural Language Models},
  author={Carlini, Nicholas and Ippolito, Daphne and Jagielski, Matthew and Lee, Katherine and Tram{\`e}r, Florian and Zhang, Chiyuan},
  booktitle={International Conference on Learning Representations},
  year={2023}
}

@misc{nasr2023scalable,
  title={Scalable Extraction of Training Data from (Production) Language Models},
  author={Nasr, Milad and Carlini, Nicholas and Hayase, Jonathan and Jagielski, Matthew and Cooper, A. Feder and Ippolito, Daphne and Choquette-Choo, Christopher A. and Wallace, Eric and Tram{\`e}r, Florian and Lee, Katherine},
  howpublished={arXiv:2311.17035},
  doi={10.48550/arXiv.2311.17035},
  year={2023}
}

@misc{zou2024poisonedrag,
  title={PoisonedRAG: Knowledge Corruption Attacks to Retrieval-Augmented Generation of Large Language Models},
  author={Zou, Wei and Geng, Runpeng and Wang, Binghui and Jia, Jinyuan},
  howpublished={arXiv:2402.07867},
  doi={10.48550/arXiv.2402.07867},
  year={2024}
}

@misc{chen2024agentpoison,
  title={AgentPoison: Red-teaming LLM Agents via Poisoning Memory or Knowledge Bases},
  author={Chen, Zhaorun and Xiang, Zhen and Xiao, Chaowei and Song, Dawn and Li, Bo},
  howpublished={arXiv:2407.12784},
  doi={10.48550/arXiv.2407.12784},
  year={2024}
}

@inproceedings{brown2020language,
  title={Language Models are Few-Shot Learners},
  author={Brown, Tom B. and Mann, Benjamin and Ryder, Nick and Subbiah, Melanie and Kaplan, Jared and Dhariwal, Prafulla and Neelakantan, Arvind and Shyam, Pranav and Sastry, Girish and Askell, Amanda and Agarwal, Sandhini and Herbert-Voss, Ariel and Krueger, Gretchen and Henighan, Tom and Child, Rewon and Ramesh, Aditya and Ziegler, Daniel M. and Wu, Jeffrey and Winter, Clemens and Hesse, Christopher and Chen, Mark and Sigler, Eric and Litwin, Mateusz and Gray, Scott and Chess, Benjamin and Clark, Jack and Berner, Christopher and McCandlish, Sam and Radford, Alec and Sutskever, Ilya and Amodei, Dario},
  booktitle={Advances in Neural Information Processing Systems},
  volume={33},
  pages={1877--1901},
  year={2020}
}

@misc{schwartzman2024exfiltration,
  title={Exfiltration of personal information from ChatGPT via prompt injection},
  author={Schwartzman, Gregory},
  howpublished={arXiv:2406.00199},
  doi={10.48550/arXiv.2406.00199},
  month={Jun},
  year={2024}
}

@misc{li2024lockpicking,
  title={Lockpicking LLMs: A Logit-Based Jailbreak Using Token-level Manipulation},
  author={Li, Yuxi and Liu, Yi and Li, Yuekang and Shi, Ling and Deng, Gelei and Chen, Shengquan and Wang, Kailong},
  howpublished={arXiv:2405.13068},
  doi={10.48550/arXiv.2405.13068},
  year={2024}
}

@misc{alizadeh2025simple,
  title={Simple Prompt Injection Attacks Can Leak Personal Data Observed by LLM Agents During Task Execution},
  author={Alizadeh, Meysam and Samei, Zeynab and Stetsenko, Daria and Gilardi, Fabrizio},
  howpublished={arXiv:2506.01055},
  doi={10.48550/arXiv.2506.01055},
  year={2025}
}

@misc{debenedetti2024agentdojo,
  title={AgentDojo: A Dynamic Environment to Evaluate Prompt Injection Attacks and Defenses for LLM Agents},
  author={Debenedetti, Edoardo and Zhang, Jie and Balunovi{\'c}, Mislav and Beurer-Kellner, Luca and Fischer, Marc and Tram{\`e}r, Florian},
  howpublished={arXiv:2406.13352},
  doi={10.48550/arXiv.2406.13352},
  year={2024}
}

@misc{reddy2025echoleak,
  title={EchoLeak: The First Real-World Zero-Click Prompt Injection Exploit in a Production LLM System},
  author={Reddy, Pavan and Gujral, Aditya Sanjay},
  howpublished={arXiv:2509.10540},
  doi={10.48550/arXiv.2509.10540},
  year={2025}
}

@misc{lan2026silent,
  title={Silent Egress: When Implicit Prompt Injection Makes LLM Agents Leak Without a Trace},
  author={Lan, Qianlong and Kaul, Anuj and Jones, Shaun and Westrum, Stephanie},
  howpublished={arXiv:2602.22450},
  doi={10.48550/arXiv.2602.22450},
  year={2026}
}

@misc{shi2025promptarmor,
  title={PromptArmor: Simple yet Effective Prompt Injection Defenses},
  author={Shi, Tianneng and Zhu, Kaijie and Wang, Zhun and Jia, Yuqi and Cai, Will and Liang, Weida and Wang, Haonan and Alzahrani, Hend and Lu, Joshua and Kawaguchi, Kenji and Alomair, Basel and Zhao, Xuandong and Wang, William Yang and Gong, Neil and Guo, Wenbo and Song, Dawn},
  howpublished={arXiv:2507.15219},
  doi={10.48550/arXiv.2507.15219},
  year={2025}
}

@misc{zhang2026agentsentry,
  title={AgentSentry: Mitigating Indirect Prompt Injection in LLM Agents via Temporal Causal Diagnostics and Context Purification},
  author={Zhang, Tian and Xu, Yiwei and Wang, Juan and Guo, Keyan and Xu, Xiaoyang and Xiao, Bowen and Guan, Quanlong and Fan, Jinlin and Liu, Jiawei and Liu, Zhiquan and Hu, Hongxin},
  howpublished={arXiv:2602.22724},
  doi={10.48550/arXiv.2602.22724},
  year={2026}
}

@misc{weng2026argus,
  title={{ARGUS}: Defending LLM Agents Against Context-Aware Prompt Injection Attacks},
  author={Weng, Shihao and Feng, Yang and Zhang, Jinrui and Xie, Xiaofei and Yu, Jiongchi and Liu, Jia},
  howpublished={arXiv:2605.03378},
  doi={10.48550/arXiv.2605.03378},
  year={2026}
}




% 	\bibitem{keshet2016prediction}
% 	Keshet, Renato, Alina Maor, and George Kour.
% 	\newblock Prediction-Based, Prioritized Market-Share Insight Extraction.
% 	\newblock In {\em Advanced Data Mining and Applications (ADMA), 2016 12th International 
%                       Conference of}, pages 81--94,2016.

% \end{thebibliography}

\end{document}